# Observation of non-trivial topological electronic structure of orthorhombic SnSe

H. J. Zheng[1,2], W. J. Shi[3,4], C. W. Wang[1,2], Y. Y. Lv[5,6], W. Xia[1,7], B. H. Li[1], F. Wu[1], S. M. He[8], K. Huang[1,2], S. T. Cui[9], C. Chen[8,10], H. F. Yang[1], A. J. Liang[1,7], M. X. Wang[1,7], Z. Sun[9], S. H. Yao[5,6], Y. B. Chen[5,6], Y. F. Guo[1,7], Q. X. Mi[1], L. X. Yang[11], M. S. Bahramy[12], Z. K. Liu[1,7,†], Y. L. Chen[1,7,8,11,†]

[1]*School of Physical Science and Technology, ShanghaiTech University, Shanghai 201210, P. R. China*
[2]*University of Chinese Academy of Sciences, Beijing 100049, P. R. China*
[3]*Center for Transformative Science, ShanghaiTech University, Shanghai 201210, P. R. China*
[4]*Shanghai high repetition rate XFEL and extreme light facility (SHINE), ShanghaiTech University, Shanghai 201210, P. R. China*
[5]*Department of Physics, Department of Materials Science and Engineering, National Laboratory of Solid State Microstructures, Nanjing University, Nanjing 210093, China*
[6]*Collaborative Innovation Center of Advanced Microstructures, Nanjing University, Nanjing 210093, China*
[7]*ShanghaiTech Laboratory for Topological Physics, Shanghai 201210, P. R. China*
[8]*Department of Physics, University of Oxford, Oxford, OX1 3PU, United Kingdom*
[9]*National Synchrotron Radiation Laboratory, University of Science and Technology of China, Hefei, Anhui 230026, China*
[10]*Advanced Light Source, Lawrence Berkeley National Laboratory, Berkeley, CA 94720, USA*
[11]*State Key Laboratory of Low Dimensional Quantum Physics, Department of Physics, Tsinghua University, Beijing 100084, China*
[12]*Department of Physics and Astronomy, The University of Manchester, Manchester M13 9PL, United Kingdom*

[†]*Email: yulin.chen@physics.ox.ac.uk, liuzhk@shanghaitech.edu.cn*

**Topological electronic structures are key to the topological classification of quantum materials and play an important role in their physical properties and applications. Recently, SnSe has attracted great research interests due to its superior thermoelectric performance. However, it's topological nature has long been ignored. In this work, by combining synchrotron-based angle-resolved photoemission spectroscopy and *ab-initio* calculations, we systematically investigated the topological electronic structure of orthorhombic SnSe. By identifying the continuous gap in the valence bands due to the**

**band inversion and the topological surface states on its (001) surface, we establish SnSe as a strong topological insulator. Furthermore, we studied the evolution of the topological electronic structure and propose the topological phase diagram in SnSe$_{1-x}$Te$_x$. Our work reveals the topological non-trivial nature of SnSe and provides new understandings of its intriguing transport properties.**

## 1. Introduction

Topological phases of matter have attracted extensive research interest in the condensed matter physics field [1–3]. Over the past decade, numerous types of topological materials have been theoretically proposed by first-principle calculation and experimentally verified by transport or spectroscopic measurements. The category now includes topological insulators [1–3], three-dimensional topological Dirac and Weyl semimetals [4–6], topological superconductors [2,7–12] and topological magnetic insulators/semimetals [13–16]. The recent rapid development of high-throughput prediction of band topology based on the theories of topological quantum chemistry [17] and symmetry-based indicators [18,19] provides us a more efficient approach to search for the topological materials with application potentials.

Similar to the well-studied topological insulators Bi$_2$Se$_3$ and Bi$_2$Te$_3$ [20–22], the Group-IV monochalcogenides family and their alloy are known as promising thermoelectric materials [23–25]. Among them, SnSe hosts the record-high thermoelectric figure-of-merit of ~2.6 at 923K [23], much higher than that of typical known high-performance thermoelectric materials [26–28]. Previous investigations on the electronic structure of SnSe have focused on its impact on the thermoelectric properties, revealed the valence bands with multiple valleys and relatively small in-plane effective masses, which are key to the enhancement of the Seebeck coefficient while keeping high electrical conductivity [29–32]. These ARPES works not only provide deeper understanding of the electronic origin of the excellent thermoelectric properties in SnSe, but also offer a guideline for improving higher ZT thermoelectric materials by band structure engineering. However, the topological aspect of SnSe has long been ignored. SnSe has two isomers: the metastable SnSe with rocksalt crystal structure (similar to SnTe) was theoretically proposed to be a topological crystalline insulator (TCI) [33], and experimentally confirmed by angle resolved photoemission spectroscopy (ARPES) in epitaxial grown films [34,35]. The thermodynamically more stable SnSe with a layered orthorhombic (Pnma, No. 62) GeS-type crystal structure [25,36] has long been considered as a topologically trivial band insulator with gap size ~0.9 eV [37]. Although a pressure-induced superconducting transition and topological phase transition was experimentally observed [38], direct evidence of topological electronic structure of SnSe has not been addressed from the previous spectroscopic works [29–32,39].

In this work, we present systematic investigation on the electronic structure of orthorhombic SnSe, by combining the use of state-of-the-art angle-resolved photoemission spectroscopy (ARPES) technique and *ab-initio* calculations. Based on the detailed theoretical analysis of the topological invariant and systematic photon energy-dependent ARPES measurement, we identified the non-trivial topological electronic structure of SnSe, including a continuous gap in the valence bands due to the band inversion and topological surface states (TSSs) within the continuous gap on its (001) surface, thus proving SnSe as a strong topological insulator (STI). Different from most other topological insulators with TSSs inside bulk gap such as $Bi_2Se_3$/$Bi_2Te_3$ [1–3,21], the TSSs of orthorhombic SnSe are buried deeply below the Fermi energy ($E_F$) and strongly overlap with the bulk valence band, demonstrating its unusual robustness. In addition, our investigation on the evolution of the band structure of SnSe$_{1-x}$Te$_x$ (x=0, 0.2 and 0.5) demonstrates the band evolution under different spin-orbit coupling (SOC) strength, where the topological non-trivial electronic structures persist. Our result further suggests a topological phase transition (STI to TCI) accompanying the structural phase transition with x>0.7 [40]. Our work deepens the understanding on the topological nature of SnSe family and sheds light on further understanding of its electronic structure and thermoelectric properties [41,42].

## 2. Methods

SnSe and SnSe$_{1-x}$Te$_x$ single crystals were grown by the vapor transport method. We first synthesize the polycrystalline samples by the direct solid-state reaction using stoichiometric quantities of Sn (99.995%, Alfa Aesar), Se (99.999%, Alfa Aesar), and Te (99.999%, Alfa Aesar). In detail, the starting materials were loaded in a preheated quartz ampoule inside the glove box, sealed under vacuum ($10^{-4}$ Pa), heated to 950 °C at 12 h, and held for 36 h to obtain polycrystalline samples. Then the amount of polycrystalline powders were mixed with iodine (I$_2$, transport agent, 2.5 mg/cm$^3$) and transferred in another sealed and evacuated ($10^{-4}$ Pa) quartz ampoule. A two-zone tube furnace was used with a temperature gradient of 100 °C between hot (source temperature, 500 °C) and cold (growth temperature, 600 °C) zones to grow crystals. Over 10 days of transport, many single crystals shaped as thin plates were obtained at the cold part of the quartz ampoule.

Synchrotron-based ARPES measurements were performed at beamline 13U of the National

Synchrotron Radiation Laboratory (NSRL), China, and beamline 03U of the Shanghai Synchrotron Radiation Facility (SSRF), China [43]. The samples were cleaved in situ and measured under ultra-high vacuum below $5\times10^{-11}$ Torr. Data were collected by Scienta DA30L analyzer. The total energy and angle resolution were below 10 meV and 0.2°, respectively.

The first-principles calculations were carried out using the Vienna ab initio Simulation Package (VASP) [44]. The valence electrons and ion cores interactions were described by the projector augmented wave (PAW) method [45,46]. The exchange-correlation potential is formulated by the generalized gradient approximation with the Perdew-Burke-Ernzerhof (PBE) scheme [47]. The Γ-centered 12×12×4 k points are used for the first Brillouin-zone sampling. The plane-wave basis cutoff energy is set to 300 eV. The tight-binding Hamiltonian was constructed using the maximally localized Wannier functions which was provided by Wannier90 package [48]. The projected atomic s and p orbitals of Sn and Se atoms were employed to reproduce the ab-initio calculated band structure. The surface states were obtained by the surface Green's function method [49]. The experiment lattice constant (Inorganic Crystal Structure Database no. 12863) was used in the calculations. In the unit cell of SnSe, there are four Sn and four Sn atoms. For $SnSe_{0.75}Te_{0.25}$, one Se atom was substituted by one Te atom, while for $SnSe_{0.5}Te_{0.5}$, two Se atoms was substituted by two Te atoms. After relaxation, the structures with lowest energy were used for the following calculation.

## 3. Results and Discussions

As shown in Figs. 1a-c, orthorhombic phase SnSe adopts a layered crystal structure of space group Pnma with inversion symmetry [36], which can be viewed as a binary analogy to the isostructural black phosphorus, an intriguing 2D material with prominent carrier mobility, thickness-dependent direct bandgap and in-plane anisotropic physical properties [50–53]. Each layer consists of a puckered honeycomb lattice of alternating Sn and Se atoms, with armchair chains along x-axis and zigzag chains along y-axis. The adjacent layers have the opposite direction along x-axis. The crystal naturally cleaves along the (001) direction, which is ideal for the ARPES measurements. The high quality of the crystals used in this work was confirmed by X-ray diffraction (details in Supplementary Information Part 1). The bulk Brillouin zone (BZ) and the projected (001) surface BZs are shown in Fig. 1d, with the high-symmetry points labeled.

    We firstly carry out detailed DFT calculations to identify the topological nature of the compound. Figs. 1e and f show the calculated bulk band structure of orthorhombic SnSe without and with SOC, respectively. In addition to the indirect bulk bandgap of about 0.9 eV [37] of SnSe, we can observe the turning on of a 'continuous gap' (shaded in gray in Figure 1f) in which there is no crossing of the bulk valence bands after the inclusion of strong SOC, together with band inversions of Sn-s and Se-p orbitals as labelled in Fig. 1f. We evaluate the topological nature of the compound based on the inversion symmetry of the crystal, by using the method proposed by Fu et al [54]. As shown in Fig.1d, the eight time-reversal-invariant-momenta (TRIM) points in the BZ are Γ, X, S, Y, Z, U, T and R, respectively. Since the other 6 TRIM points have the same parities, the $Z_2$ invariant $\nu_0$ can be calculated by considering parities of Γ and T points solely, resulting in $\nu_0=1$. The result suggests the orthorhombic SnSe is an STI which has been ignored in previous reports [29–32,39]. Further, we explicitly unveil the characteristic TSS in SnSe by calculating the spectral function of the (001) surface using the semi-infinite surface model (details in SI). As shown in Fig.2a, multiple non-trivial TSSs have

been predicted inside the continuous gap of SnSe, as highlighted by the red color. The most prominent one could be found around the $\bar{X}$ point along the $\bar{\Gamma}-\bar{X}$ direction.

We verify the predicted topologically non-trivial electronic structure by performing synchrotron-based ARPES measurements. The overall measured electronic structure of SnSe on the (001) surface is illustrated in Fig. 2b. The photoemission spectra are recorded by using hν = 21 eV photons, roughly corresponding to the $k_z$=0 case (The detailed relationship between photon energies and $k_z$ have been reported in our previous work [30]). Fig. 2b presents the measured electronic band structure in the 3D volume plot containing both the $\bar{\Gamma}$ point and $\bar{X}$ point at BZ boundary, showing the overall agreement with the *ab-initio* calculations (the detailed comparison is also presented in ref [30]).

As shown in Figs. 2d(i) and e(i), in order to examine the topological character of the electronic structures, we zoom in the band dispersions along the high symmetry $\bar{\Gamma}-\bar{X}$ and $\bar{X}-\bar{S}$ directions. The major features are contributed by the four topmost bulk valence bands α, $β_1$, $β_2$ and γ, and an obvious SOC-induced gap between $β_1$ and $β_2$ was observed in Fig. 2d(i), which is predicted to be a band-crossing at ~0.1Å$^{-1}$ away from $\bar{X}$ along the $\bar{\Gamma}-\bar{X}$ direction without SOC (Fig. 1c). Interestingly, the α and $β_1$ bands, as well as $β_2$ and γ bands are degenerate at $\bar{X}$ point. According to the *ab-initio* calculation, the degenerations along the $\bar{X}-\bar{S}$ line without SOC (Fig.1c) are lifted with SOC (Fig. 1d and 2e(ii)). Unfortunately, these degeneration lifting cannot be clearly distinguished within our energy resolution, possibly due to the strong $k_z$ broadening effect [55]. We further address the topological surface state by making a detailed comparison of the measured band dispersions and the semi-infinite surface calculations in Fig.2d(iii) and e(iii), and observe an excellent agreement. The TSSs near the $\bar{X}$ predicted by the calculation were clearly seen in experimental band dispersions as in gap (i.e., the continuous gap) states connecting the $\beta_1$ and $\beta_2$ bands. They appear as a flat 'X' shape along the $\overline{\Gamma}-\overline{X}$ direction (Fig. 2d) and much dispersive along the $\overline{X}-\overline{S}$ direction between the $\beta_1$ and $\beta_2$ bands. The existence of the TSSs can also be evidenced in the constant energy contours at the 0.4 eV below the Fermi level (Fig. 2c). The dispersion of another TSS could be observed at a deeper binding energy in the ARPES measurement along the $\overline{X}-\overline{S}$ direction (Fig. 2e).

We probe the surface nature of the observed TSSs by photon-energy-dependent ARPES measurements using photon energies from 19 to 29 eV, covering more than half of the bulk BZ [30]. As shown in Fig. 3, the TSSs observed along both $\overline{\Gamma}-\overline{Y}$ and $\overline{Y}-\overline{S}$ directions show little variations in the dispersion under all photon energies, clearly demonstrating the absence of $k_\perp$ dispersion of the TSSs.

We further investigate the evolution of the topological nontrivial electronic structure of $SnSe_{1-x}Te_x$ with Te substituting Se. $SnSe_{1-x}Te_x$ remains a stable orthorhombic crystal structure up to x=0.5 before the structural transition to the rocksalt phase [40]. In Fig.4a, the x-ray photoelectron spectra (XPS) of three x values ($SnSe$, $SnSe_{0.8}Te_{0.2}$ and $SnSe_{0.5}Te_{0.5}$) clearly show characteristic core level peaks of Se, Te and Sn, which suggest their high crystal quality and differences in chemical compositions, respectively. With the increase of Te composition, as evidenced by the growing peak of Te 3d doublets (the zoom-in plot of Fig. 4 (a)), the SOC strength should be enhanced. With the stronger SOC and Te substitution the electronic structure of $SnSe_{1-x}Te_x$ experiences complex evolution, the continuous gap deforms but still remains and the TSS survives, as predicted by the semi-infinite surface calculations (Figs. 4e-g). Our ARPES measurement of $SnSe_{0.8}Te_{0.2}$ and $SnSe_{0.5}Te_{0.5}$ nicely repeats the predicted band structure

(Figs. 4b-d), where the observed inner electron pocket of topmost valence band moves to deeper binding energy comparing to the SnSe case. The TSS deeply buries into the bulk continuum and could no longer be distinguished. Our calculation and ARPES measurement, along with previously reported TCI in SnTe [56], suggest the SnSe$_{1-x}$Te$_x$ hosts a strong topological insulator phase across wide x range before experiencing the structural [40] and topological phase transition into the rocksalt/topological crystalline insulator phase (Fig. 4d).

Our observation of the unusual TSSs in orthorhombic phase SnSe establish it as the first example with non-trivial topological electronic structure in the Group-IV monochalcogenides family crystallizing into the thermal-stable orthorhombic phase, instead of metastable rocksalt phase. The *ab-initio* calculation and following parity criterion based $Z_2$ invariant analysis accurately identify it as a strong topological insulator. The deeply buried TSSs show unusual robustness as they strongly overlap with bulk bands and survives under Te substitution, which suggest the possibilities to utilize the TSS in potential electronic applications.

The result on SnSe$_{1-x}$Te$_x$ also suggests great tunability in band structure engineering, providing a great platform to understand the interplay between the crystalline symmetry, SOC strength, electronic structure, thermoelectric property [41,42,57]. The rich topological phenomena in the group-IV monochalcogenides family makes them interesting playground for the realization of novel topological effects and topological phases. In addition, the electronic structure and related topological nature near the critical point of the structural and topological phase transition in SnSe$_{1-x}$Te$_x$ merit further investigations.

**Figures:**

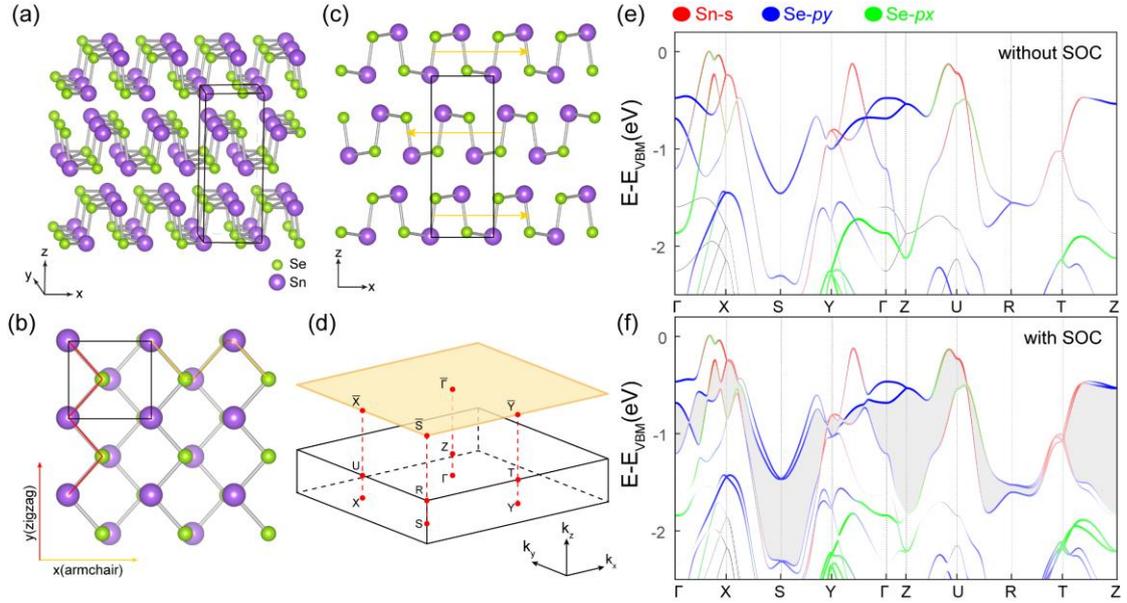

**Figure 1. Crystal structure and bulk band structure calculation. (a)** 3D view of layered orthorhombic crystal structure of thermodynamically stable orthorhombic phase SnSe. **(b)** Top view of a SnSe layer, zigzag and armchair chains are labeled by red and yellow lines. **(c)** Side view perpendicular to the armchair direction, showing two adjacent layers have the opposite direction along x-axis **(d)** The corresponding 3D Brillouin zone (BZ) of SnSe and its projected surface BZ to (001) surface, with high symmetry points labeled. **(e, f)** First-principles band calculations without and with spin-orbit coupling (SOC), respectively. Gray shaded area in **f** shows the continues gap induced by SOC. Red, blue and green dots indicate the Sn-s, Se-py and Se-px orbital contributions, respectively.

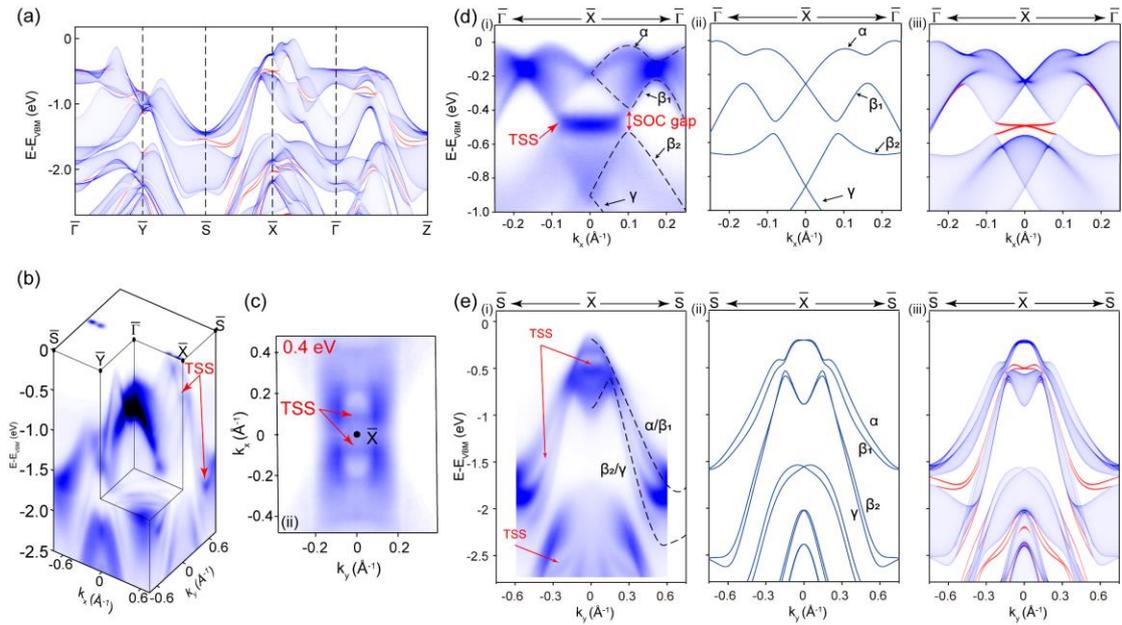

**Figure 2. Observation of Topological Surface States on SnSe (001) surface. (a)** The calculation result shown along the high-symmetry lines of the surface Brillouin zone.

Topological surface states are clearly visible and highlighted by red color. **(b)** A 3D intensity plot of the photoemission spectra around $\bar{X}$ point with TSS indicated by red arrows. **(c)** Spectral intensity plot showing constant energy contours around $\bar{X}$ point at $E-E_{VBM} = 0.4$ eV. **(d,e)** Photoemission intensity(left panel), corresponding ab initio bulk(middle panel) and semi-infinite surface(right panel) calculation along the high symmetry $\bar{\Gamma}-\bar{X}$(d), $\bar{S}-\bar{X}$ (e) directions. The TSSs are indicated by the red arrows in (i) and highlighted by red color in (iii).

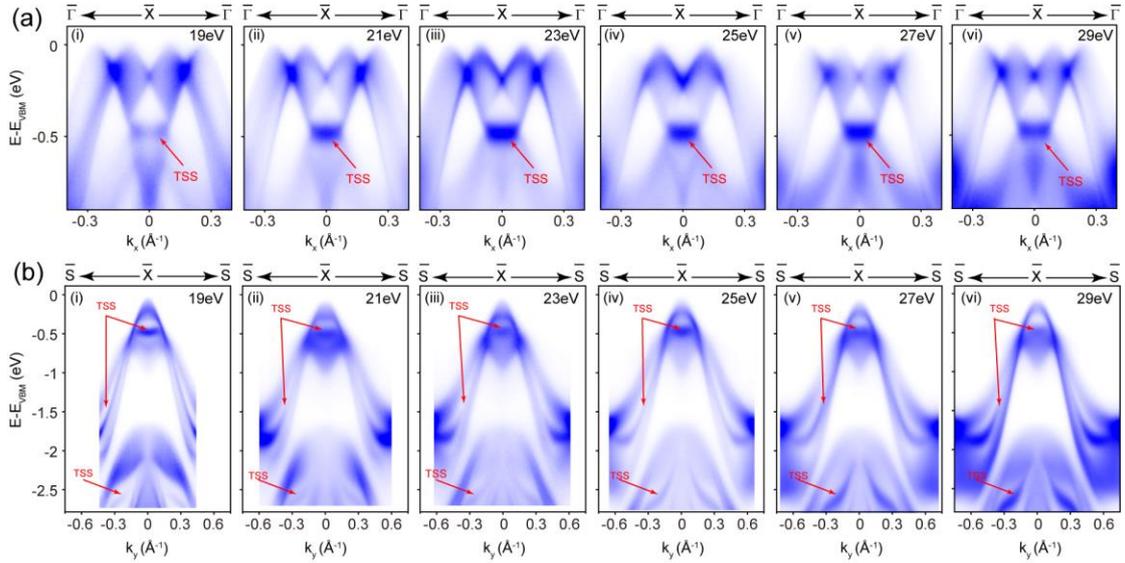

**Figure 3. Photon energy dependent of the TSS on SnSe (001) surface. (a,b)** Photoemission intensity plot along the high-symmetry $\bar{\Gamma}-\bar{X}$ (a) and $\bar{S}-\bar{X}$ (b) directions with photon energies from 19 to 29 eV covering more than half bulk BZ. Red arrows indicate TSSs that do not change with the photon energies, although the transition matrix element effects modulate the intensity shape of TSS captured by different photons.

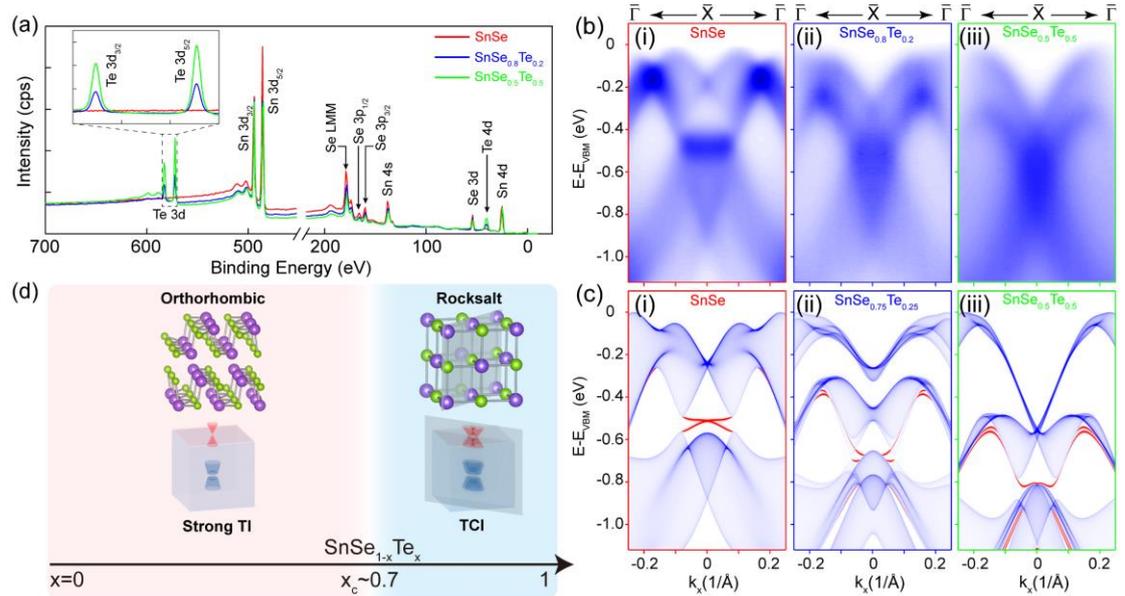

**Figure 4. Evolution of the band structure of SnSe with different tellurium doping amounts.** (a) The x-ray photoelectron spectra (XPS) of SnSe, SnSe$_{0.8}$Te$_{0.2}$, SnSe$_{0.5}$Te$_{0.5}$ clearly shows their characteristic core level peaks. Zoom-in plot of Te 3d doublets indicate different contents of Te element in three compounds. (b) and (c) Comparison of measured band dispersions (upper panel) and corresponding calculated band structures (lower panel) around $\bar{X}$ point for SnSe, SnSe$_{0.8}$Te$_{0.2}$, SnSe$_{0.5}$Te$_{0.5}$, respectively. The TSSs are highlighted by red color in (c). (d) Sketch of a structural and topological phase transition across a critical point x$_c$ in SnSe$_{1-x}$Te$_x$.